\documentclass{intech}

\usepackage[latin1]{inputenc}
\usepackage{colortbl}
\usepackage[english]{babel}

\booktitle{Advances in Spacecraft Technologies} 

\chaptertitle{Nonlinear Electrodynamics: Alternative Field Theory for Featuring 
Photon Propagation Over Weak Background Electromagnetic Fields and what Earth 
Receivers Read off Radio Signals from Interplanetary Spacecraft Transponders } 

\authors{Herman J. Mosquera Cuesta$^{1,2,3}$}
\affiliation{
$^1$Departmento de F{\'\i}sica Universidade Estadual Vale do Acara\'u
\\Avenida da Universidade 850, Campus da Bet\^ania, CEP 62.040-370, Sobral, 
Cear\'a\\
$^2$Instituto de Cosmologia, Relatividade e Astrof\'\i sica (ICRA-BR)\\
Centro Brasileiro de Pesquisas F{\'\i}sicas, Rua Dr. Xavier Sigaud 150,\\
CEP 22290-180, Urca Rio de Janeiro, RJ\\
$^3$International Center for Relativistic Astrophysics Network (ICRANet)\\
International Coordinating Center, Piazzalle della Repubblica 10, 065112, Pescara}
\country{$^{1,2}$Brazil\\$^3$Italy}

\begin{document}


\newcommand*{\be}{\begin{equation}}
\newcommand*{\ee}{\end{equation}}

\maketitle

\section{Introduction}

A few observational and/or experimental results have dramatically pushed forward 
the research program on gravity as those from the radio-metric Doppler tracking 
received from the Pioneer 10 and 11 spacecrafts when the space vehicles were at 
heliocentric distances between 20 and 70 Astronomical Units (AU). These data have 
conclusively demonstrated the presence 
of an anomalous, tiny and blue-shifted frequency drift that changes smoothly at a rate of $ 
\sim 6 \times 10^{-9}$ Hz s$^{-1}$. Those signals, if interpreted as a gravitational pull 
of the Sun on each Pioneer vehicle, translates into a deceleration of $a_P = (8.74\pm 
1.33) \times 10^{-10}$ m s$^{-2}$. This sunward acceleration appears to be a violation
of Newton's inverse-square law of gravitation, and is referred to as the Pioneer 
anomaly, the nature of which remains still elusive to unveil. 

Within the theoretical framework of nonlinear electrodynamics (NLED) in what 
follows we will address this astrodynamical puzzle, which over the last fifteen 
years has challenged in a
fundamental basis our understanding of gravitational physics. To this goal we 
will first, and briefly, review the history of the Pioneers 10 and 11 missions.
Then a synopsis of currently available Lagrangian formulations of NLED is given. 
And  finally, we present our solution of this enigma by invoking a special class of 
NLED theories featuring a proper description of electromagnetic phenomena taking 
place in environments where the strength of the (electro)magnetic fields in the background 
is decidedly low.

\section{What is the problem: The Pioneer anomaly}

In this short voyage to the Pioneer 10 and 11 missions our main guide will be 
the comprehensive and richly documented recent review on the Pioneer Anomaly 
by [Turyshev, S. G.  \& Toth, V. T. (2010).  Living Rev. Rel. 13 (2010) 4. 
arXiv:1001.3686, v2, gr-qc] from which we retake some ideas
and references. (The attentive readers are kinldy addressed to this invaluable 
article).

The Pioneer 10 and 11 spacecrafts were the first two man-made space 
vehicles designed to explore the outer solar system. The trajectories of the 
spaceships were projected to passage nearby Jupiter during 1972-1973 
having as objectives to conduct exploratory investigation of the 
interplanetary medium beyond the orbit of Mars, the nature of the 
asteroid belt, the environmental and atmospheric characteristics of 
Jupiter and Saturn (for Pioneer 11), and to investigate the solar 
system beyond the orbit of the Jovian planet.\footnote{See details 
on the Pioneer missions at 
\url{http://www.nasa.gov/centers/ames/missions/archive/pioneer.html}.
Be awared that another member of Pioneer spacecrafts family, Pioneer 
6, remained operational for more than 35 years after launch.}

The Pioneer missions were the first space probes to adventure over the 
asteroid belt, heading for close-up observations of the gaseous giant 
planets, and for performing in situ studies of the physical properties 
of the interplanetary medium in the outer solar system. The design of their
missions was guided by the simplicity, having a powerful rocket-launching 
system to push the spacecrafts on an hyperbolic trajectory aimed directly at 
Jupiter, which the spacecrafts were expected to fly-by approximately 21 months 
after launch (see Fig. 1).

\begin{figure}[t!]
\centering
\includegraphics[height=10cm]{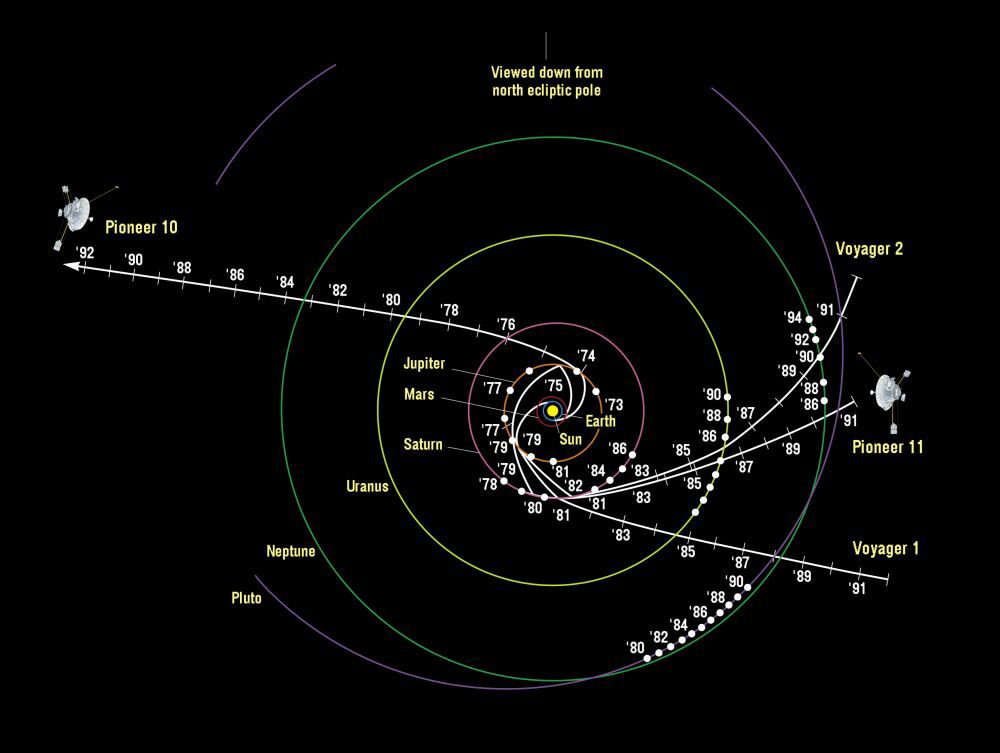}
\caption{Ecliptic pole view of the spacecrafts Pioneer 10 
and 11 interplanetary trajectories (see also the trajectories 
of the vehicles Voyager 1 and 2). Credit: 
\url{http://www.nasa.gov/centers/ames/missions/archive/pioneer.html}}
\label{pioneers}
\vspace{-12pt}
\end{figure}

By the late 1960's, the aerospace engineering technology available to the 
designers of the Pioneer missions made it no longer practical to use solar 
panels for operating a spacecraft at large distances, as for instance that 
of Jupiter. A cause of this, a built-in nuclear power system, in the form 
of radioisotope 
thermoelectric generators (RTGs) powered by $^{238}$Pu, was chosen as the 
means to provide electrical power to the spaceship. As even this was 
relatively new technology at the time the missions were designed, the power 
subsystem was suitably over-engineered, being the unique design requirement 
to have a completely functional space probe capable of performing all planned 
scientific tasks by running only three (out of four) RTGs.

The entire design of these spacecrafts and their science missions was 
characterized by such conservative engineering, and for sure it was 
responsible for both the exceptional longevity of the two spacecrafts 
and their ability to deliver science results which by far exceeded the 
expectations of their designers. 

The original plan envisioned a primary 
mission of two to three years in duration. Nevertheless, following its 
encounter with Jupiter, Pioneer 10 remained functional for over 30 years. 
Meanwhile, Pioneer 11, though not as long lived as its engineering-copy 
craft, successfully navigated a path across the solar system for another
encounter with Saturn, offering the first close-up observations of 
the ringed planet. After the encounters with Jupiter and Saturn (for 
Pioneer 11, see Fig. 1), the space ships followed, near the plane of the 
ecliptic, hyperbolic orbits of escape heading to opposite sides of 
the solar system, continuing their extended missions. The spacecrafts 
explored the outer regions of the solar system, studying energetic 
particles from the Sun (solar wind), and cosmic rays entering our 
neighborhood in the Milky Way. (Their cousin spacecrafts, the Voyager1 
and 2, that where launched contemporarily, studied in the beginning 
of their mission, the interplanetary space, what resulted in a very 
accurate mapping of the interplanetary magnetic field and its strength, 
as one can see in Fig. 2 below).

In virtue of a combination of many factors, the Pioneers were excellent space 
sondes for pursuing experiments of high precision celestial mechanics. This 
includes the presence of a coherent mode transceiver on board, the attitude 
control (spin-stabilized, with a minimum number of attitude correction 
maneuvers using thrusters), the design of the power system (the RTGs being 
on extended booms aided the stability of the craft and also helped in reducing 
thermal effects), and Doppler tracking of high precision (with the accuracy 
of post-fit Doppler residuals at the level of mHz). The exceptional built-in
sensitivity to acceleration of the Pioneer 10 and 11 spacecrafts naturally 
allowed them to reach a level of accuracy of $\sim 10^{-10}$ m/s$^2$. The 
result was one of the most precise spacecraft navigations in deep space 
since the early days of space exploration. That is the great legacy of the 
Pioneer missions.

After having had a brief accounting of the Pioneers missions, one can proceed to
review our current understanding of nonlinear electrodynamics and to settle down
the foundations for its use in the search for a solution to the Pioneer anomaly. In 
this Section we shall briefly review the theoretical foundations of some theories 
of  NLED, focusing essentially on the fundamental prediction concerning the way 
photons propagate through a vacuum space permeated by electromagnetic (EM) fields: 
The fact that photons travel along the \textit{effective metric}, and not over the 
geometry in the background. It is this peculiar feature what makes the photon to 
``feel'' itself being acted upon by a force, and consequently to undergo acceleration.
\footnote{Because of the special theory of relativity constraints regarding the 
propagation of any perturbation, it becomes clear that such effect must manifest 
itself as a change in one or both of their physical properties: its frequency or its 
wavelength. Hence, through the Pioneer spacecrafts radio Doppler tracking we 
might be observing the effect on the photon frequency.} In our understanding, such 
effect is responsible for the drift in frequency undergone by the photon. Next we 
will show that any NLED, independently of the specific form of its Lagrangian, 
brings in such a frequency shift. And in our view, it is such acceleration what 
can account for the Pioneer anomaly.

\section{Some Lagrangian formulations of nonlinear electrodynamics}

To start with, it is worth to recall that 
according to quantum electrodynamics (QED: see \cite{delphenich2003, delphenich2006} for a 
complete review on NLED and QED) a vacuum has nonlinear properties
(Heisenberg \& Euler 1936; Schwinger 1951) which affect the
 photon propagation.  A noticeable advance in the realization of 
this theoretical prediction has been provided by [Burke, Field,  
Horton-Smith , etal., 1997), who  demonstrated experimentally 
that the inelastic scattering of laser photons by gamma-rays in a 
background magnetic ield  is definitely a nonlinear 
phenomenon. The propagation of photons in NLED has been examined by 
several authors [Bialynicka-Birula \& Bialynicki-Birula, 1970; Garcia 
\& Plebanski, 1989; Dittrich \& Gies, 1998; De Lorenci, Klippert, 
Novello, etal., 2000; Denisov, Denisova \&  Svertilov, 2001a, 2001b, 
Denisov \& Svertilov, 2003]. In the geometric optics approximation, it 
was shown by [Novello, De Lorenci, Salim \& etal., 2000;  
Novello \& Salim, 2001], that when the photon propagation is 
identified with the propagation of discontinuities of the EM field in a
nonlinear regime, a remarkable feature appears:  The discontinuities
propagate along null geodesics of an {\it effective} geometry which 
depends on the EM field on the background. This means that the NLED 
interaction can be geometrized. An immediate consequence of this 
NLED property is the prediction of the phenomenon dubbed as photon 
acceleration, which is nothing else than a shift in the frequency of 
any photon traveling over background electromagnetic fields. 
The consequences of this formalism are examined next.

\subsection{Heisenberg-Euler approach}

The Heisenberg-Euler Lagrangian for nonlinear electrodynamics 
 (up to order 2 in the truncated infinite series of terms involving $F$) 
has the form \cite{heisenberg-euler36} 

\be
L_{\rm H-E} = -\frac{1}{4} F + \bar{\alpha} F^2 + \bar{\beta} G^2 \, ,
\ee

where $F = F_{\mu \nu} F^{\mu \nu}$, with $F_{\mu \nu} = \partial_\mu 
A_\nu - \partial_\nu A_\mu$, and $G = \frac{1}{2}\eta_{\alpha 
\beta \gamma \delta} F^{\alpha \beta} F^{\gamma \delta} = -4 \vec{E} 
\cdot \vec{B}$, with greek index running (0, 1, 2, 3), while $\bar{
\alpha}$ and $\bar{\beta}$ are arbitrary constants. 

When this Lagrangian is used to describe the photon dynamics the 
equations for the EM field in vacuum coincide in their form with the 
equations for a continuum medium in which the electric permittivity 
and magnetic permeability tensors $\epsilon_{\alpha\beta}$ and $\mu_{\alpha \beta}$ 
are functions of the electric and magnetic fields determined by
some observer represented by its 4-vector velocity $V^{\mu}$ [Denisov,
Denisova \& Svertilov, 2001a, 2001b; Denisov \& Svertilov, 2003; Mosquera 
Cuesta \& Salim, 2004a, 2004b]. The attentive reader must notice that this 
first order approximation is valid only for $B$-fields smaller than 
$B_q = \frac{m^2 c^3}{e \bar{h}} = 4.41\times10^{13}$~G (Schwinger's 
critical $B$-field \cite{Schwinger51}). In curved spacetime, these equations 
are written as 

\begin{equation}
D^{\alpha}_{||\alpha}=0, \hskip 0.5 truecm B^{\alpha}_{||\alpha}=0\; ,
\end{equation}


\begin{equation}
D^{\alpha}_{||\beta} \frac{V^{\beta}}{c} +
\eta^{\alpha \beta \rho\sigma} V_{\rho} H_{\sigma||\beta}=0,
\end{equation}

\begin{equation}
B^{\alpha}_{||\beta}\frac{V^{\beta}}{c} -
\eta^{\alpha\beta\rho\sigma}V_{\rho}E_{\sigma||\beta}=0\, .
\end{equation}

Here,  the  vertical bars subscript  ``$_{||}$'' stands for covariant 
derivative and $\eta^{\alpha\beta\rho\sigma}$ is the antisymmetric 
Levi-Civita tensor. 

The 4-vectors representing the electric and magnetic fields are defined as usual in terms of 
the electric and magnetic fields tensor $F_{\mu\nu}$ and polarization tensor $P_{\mu\nu}$

\begin{equation}
E_{\mu}=F_{\mu\nu}\frac{V^{\nu}}{c}, \hskip 0.5 truecm B_{\mu} = 
F^{*}_{\mu\nu}\frac{V^{\nu}}{c}\; ,
\end{equation}

\begin{equation}
D_{\mu}=P_{\mu\nu}\frac{V^{\nu}}{c}, \hskip 0.5 truecm H_{\mu} = 
P^{*}_{\mu\nu}\frac{V^{\nu}}{c}\; ,
\end{equation}

where the dual tensor $X^{*}_{\mu\nu}$ is defined as $ X^{*}_{\mu\nu} 
= \frac{1}{2} \eta_{\mu\nu \alpha\beta} X^{\alpha\beta}\;$,
for any antisymmetric second-order tensor $X_{\alpha\beta}$.

The meaning of the vectors $D^{\mu}$ and $H^{\mu}$ comes from the
Lagrangian of the EM field,  and in the vacuum case they  are given by   

\begin{equation}
H_{\mu}= \mu_{\mu\nu}B^{\nu},  \hskip 0.5 truecm D_{\mu}=\epsilon_{\mu\nu}
E^{\nu}\; ,
\end{equation}

where the permeability and tensors are given as

\begin{equation}
\mu_{\mu\nu}= \left[1+ \frac{2 \alpha}{45 \pi B^2_q}\left(B^2-E^2\right)
\right]h_{\mu\nu} - \frac{7 \alpha}{45 \pi B^2_q} E_{\mu} E_{\nu}\; ,
\label{permittivity-mu}
\end{equation}

\begin{equation}
\epsilon_{\mu\nu} =\left[1+ \frac{2 \alpha}{45 \pi B^2_q}\left(B^2-E^2\right)
\right]h_{\mu\nu} + \frac{7 \alpha}{45 \pi B^2_q} B_{\mu} B_{\nu}\; .
\label{permittivity-epsilon}
\end{equation}

In these expressions $\alpha$ is the EM coupling constant $(\alpha =
\frac{e^2}{\hbar c} = \frac{1}{137})$. The tensor $h_{\mu\nu}$ is the 
metric induced in the reference frame perpendicular to the observers
determined by the vector field $V^\mu$.

Meanwhile, as we are assuming that $E^\alpha = 0$, then one gets 

\be
\epsilon^\alpha_\beta = \epsilon h^\alpha_\beta + \frac{7 \alpha} 
{45 \pi B^2_q} B^\alpha B_\beta 
\ee

and $\mu_{\alpha\beta} = \mu 
h_{\alpha\beta}$. The scalars $\epsilon$ and $\mu$ can be read 
directly from Eqs.(\ref{permittivity-mu}, \ref{permittivity-epsilon}) 
as 

\be
 \epsilon \equiv \mu = 1 + \frac{2 \alpha}{45 \pi B^2_q}B^2 \, .
\ee

Applying conditions (\ref{eq14}) and (\ref{discontinuity}) (derived in  
the Appendix) to the field equations when $E^\alpha = 0$, we obtain the
constraints $e^{\mu} \epsilon_{\mu\nu} k^{\nu} = 0$ and $b^{\mu} k_{\mu}
= 0$ and the following equations for the discontinuity fields
$e_\alpha$ and $b_\alpha$:

 \begin{equation}
 \epsilon^{\lambda\gamma}e_{\gamma}k_{\alpha} \frac{V^\alpha}{c} + 
\eta^{\lambda\mu\rho\nu} \frac{V_\rho}{c} \left(\mu b_\nu k_\mu - 
{\mu^\prime} \lambda_\alpha B_\nu k_\mu \right)=0 \; , 
\label{dispersao-1}
 \end{equation}

 \begin{equation}
 b^{\lambda}k_{\alpha}\frac{V^\alpha}{c} - \eta^{\lambda\mu\rho\nu} 
\frac{V_\rho}{c} \left(e_\nu  k_\mu \right)=0 \; .  
\label{dispersao-2}
 \end{equation}

Isolating the discontinuity field from (\ref{dispersao-1}), substituting
in equation (\ref{dispersao-2}), and expressing the products of the
completely anti-symmetric tensors $\eta_{\nu\xi\gamma\beta} \eta^{\lambda
\alpha\rho\mu}$ in terms of delta functions \cite{stephani2004}, we obtain

\begin{eqnarray}
&& b^{\lambda}(k_\alpha k^\alpha)^2 + \left(\frac{\mu'}{\mu}l_\beta b^\beta k_\alpha 
B^\alpha + \frac{\beta B_\beta b^\beta B_\alpha k^\alpha} {\mu-\beta B^2}\right) 
k^{\lambda} + \nonumber \\
&& \left(\frac{\mu'}{\mu l_\alpha b^\alpha} (k_\beta V^\beta)^2 
(k_\alpha k^\alpha)^2 - \frac{\beta B_\alpha b^\alpha (k_\beta 
k^\beta)^2}{\mu-\beta B^2}\right) B^{\lambda} - \left(\frac{\mu'}
{\mu} l_\mu b^\mu k_\alpha  B^\alpha k_\beta V^\beta\right)V^{\lambda} = 0 \; . 
\label{dispersion-01} 
\end{eqnarray}

This expression is already squared in $k_\mu$ but still has an unknown
$b_\alpha$ term. To get rid of it, one multiplies by $B_\lambda$, to
take advantage of the EM wave polarization dependence.  By noting that
if $B^\alpha b_\alpha = 0$ one obtains the {\it dispersion relation} by
separating out the $k^{\mu} k^{\nu}$ term, what remains is the (-)
effective metric. Similarly, if $B_\alpha b^\alpha \neq 0$, one simply
divides by $B_\gamma b^\gamma$ so that by factoring out $k^{\mu} k^{\nu}$, 
what results is the (+) effective metric. For the case $ B_\alpha
b^\alpha = 0$, one obtains the standard dispersion relation

 \begin{equation}
  g^{\alpha\beta} k_\alpha k_\beta = 0 \; .
 \end{equation}
 
 whereas for the case $ B_\alpha b^\alpha \neq 0$, the result is

 \begin{eqnarray}
 \left[\left(1+\frac{\mu'B}{\mu} + \frac{\tilde{\beta} B^2}{\mu-\tilde{\beta} B^2}\right)
  g^{\alpha\beta} - \frac{\mu'B}{\mu}\frac{V^\alpha V^\beta}
  {c^2} + \left(\frac{\mu'B}{\mu} \right. \right.
 + \left. \left. \frac{\tilde{\beta} B^2}{\mu-\tilde{\beta} B^2}\right) l^{\alpha}l^{\beta}
\right] k_{\alpha} k_{\beta} = 0 \; , 
\label{general-dynamics}
  \end{eqnarray}

where $(')$ stands for $\frac{d}{dB}$, and we have defined 

\be
\tilde{\beta} = \frac{7 \alpha} {45 \pi B^2_q} , \hskip 1.0truecm  \textrm{and}  
\hskip 1.0truecm l^\mu \equiv  \frac{B^\mu}{|B^\gamma B_\gamma|^{1/2}}
\ee

 as the unit 4-vector along the $B$-field direction.

From the above expressions we can read the effective metric $g^{\alpha
\beta}_{+}$ and $g^{\alpha\beta}_{-}$, where the labels ``+'' and ``-''
refers to extraordinary and ordinary polarized rays, respectively. Then, 
we need the covariant form of the metric tensor, which is obtained from 
the expression defining the inverse metric $ g_{\mu\nu} g^{\nu\alpha} = 
\delta^{\alpha}_{\mu }$. So that one gets from one side

\begin{equation}
g^{-}_{\mu\nu} = g_{\mu\nu} \; \label{g-}
\end{equation}

and from the other

  \begin{eqnarray}
g^{+}_{\mu\nu} = \left(1+\frac{\mu'B}{\mu} + \frac{\beta
B^2}{\mu-\beta B^2}\right)^{-1} g_{\mu\nu} \nonumber \\
 + \left[\frac{\mu'B}{\mu (1+\frac{\mu'B}{\mu}+ \frac{\beta
B^2}{\mu-\beta B^2})(1+ \frac{\beta B^2}{\mu-\beta B^2})}
\right] \frac{V_{\mu}V_{\nu}}{c^2} + \left(\frac{\frac{\mu'B}{\mu}
+ \frac{\beta B^2}{\mu-\beta B^2}}{1+\frac{\mu'B}{\mu} + 
\frac{\beta B^2}{\mu-\beta B^2}} \right) \; \; l_\mu l_\nu \; . 
\label{l-dependent}
  \end{eqnarray}
 
The function $\frac{\mu'B}{\mu}$ can be expressed in terms of the 
magnetic permeability of the vacuum, and is given as 

\be
\frac{\mu'B} {\mu} = 2\left(1-\frac{1}{\mu}\right)\, .
\ee
 
Thus equation (\ref{l-dependent}) indicates that the photon propagates 
on an effective metric.

\subsection{Born-Infeld theory}

The propagation of light can also be viewed within the framework of the 
Born-Infeld Lagrangian. Such theory is inspired in the special theory of 
relativity, and indeed it incorporates the principle of relativity in its 
construction, since the fact that nothing can travel faster than light in a 
vacuum is used as a guide to establishing the existence of an upper 
limit for the strength of electric fields around an isolated charge,  an 
electron for instance. Such charge is then forced to have a  characteristic size 
\cite{born-infeld34}. The Lagrangian then reads

\be 
L  = - \frac{b^2}{2} \left[ \left(1 + \frac{F}{ b^2} \right)^{1/2} - 1 
\right] \; . 
\label{B-I-lagrangean} 
\ee

As in this particular case, the Lagrangian is a functional of the invariant 
$F$, i.e., $L = L(F)$,
but not of the invariant $G \equiv B_\mu E^\mu$, the study of the NLED 
effects turns out to be simpler (here again we suppose $E$ = 0). In the 
equation above, $b = \frac{e}{R_0^2} = \frac{e}{\frac{e^4}{m_0^2 c^8}} = 
\frac{m_0^2 c^8}{e^3} = 9.8 \times 10^{15}\; {\rm e.s.u.}$ 

In order to derive the effective metric that can be deduced from the B-I 
Lagrangian, one has therefore to work out, as suggested in the Appendix, 
the derivatives of the Lagrangian with respect to the invariant $F$. The 
first and second derivatives then reads

\be 
L_{\rm F} = \frac{-1}{4 \left(1 + \frac{F}{ b^2}\right)^{1/2}
} \hskip 1.0truecm {\rm and} \hskip 1.0truecm L_{\rm FF} = 
\frac{1}{8 b^2 \left(1 + \frac{F}{
b^2}\right)^{3/2} } \; . 
\label{B-I-derivate1} 
\ee

The $L(F)$ B-I Lagrangian produces, according to Eq.(\ref{eff-metric}) 
in the Appendix, an {\it effective} contravariant metric given as

\be 
g^{\mu \nu}_{\rm eff}  = \frac{-1}{4 \left(1 + \frac{F}{
b^2}\right)^{1/2} } g^{\mu \nu} + \frac{B^2}{2 b^2 \left(1 +
\frac{F}{ b^2}\right)^{3/2} } \left[ h^{\mu \nu} + l^\mu l^\nu
\right] \; . 
\ee

Both the tensor $h_{\mu\nu}$ and the vector $l^\mu$ in this equation were 
defined earlier (see Eqs.(\ref{permittivity-epsilon}) and (\ref{general-dynamics}) 
above).

Because the geodesic equation of the discontinuity (that defines
the effective metric, see the Appendix) is conformal invariant, 
one can multiply this last equation by the conformal factor ${-4} 
\left(1 + \frac{F}{ b^2}\right)^{3/2}$ to obtain

\be 
g^{\mu \nu}_{\rm eff}  = \left(1 + \frac{F}{ b^2}\right)
g^{\mu \nu} - \frac{2 B^2}{b^2} \left[ h^{\mu \nu} + l^\mu l^\nu
\right] \; . 
\ee

Then, by noting that

\be 
F =  F_{\mu \nu} F^{\mu \nu} = - 2 (E^2 - B^2) \; ,
\ee

and recalling our assumption $E = 0$, then one obtains $F = 2 B^2$. 
Therefore, the effective metric reads

\be 
g^{\mu \nu}_{\rm eff}  = \left(1 + \frac{2 B^2}{ b^2}\right) g^{\mu \nu}
-\frac{2 B^2}{ b^2} \left[ h^{\mu \nu} + l^\mu l^\nu \right] \; ,
\ee

or equivalently

\be 
g^{\mu \nu}_{\rm eff}  =  g^{\mu \nu}  + \frac{2 B^2}{ b^2}
V^\mu V^\nu - \frac{2 B^2}{ b^2} l^\mu l^\nu \; . 
\ee

As one can check, this effective metric is a functional of the
background metric $g^{\mu \nu}$, the 4-vector velocity field of 
the inertial observers $V^\nu$, and the spatial configuration 
(orientation $l^\mu$) and strength of the $B$-field.

Thus the covariant form of the background metric can be obtained 
by computing the inverse of the effective metric $g^{\mu \nu}_{\rm 
eff}$ just derived. With the definition of the inverse metric $ 
g^{\mu \nu}_{\rm eff} g_{\nu \alpha}^{\rm eff} = \delta^{\mu}_{{\;} 
{\;} \alpha}$, the covariant form of the effective metric then reads 

\be 
g_{\mu \nu}^{\rm eff} =  g_{\mu \nu} - \frac{2B^2/b^2}{
(2B^2/b^2 + 1) } V_\mu V_\nu  + \frac{2B^2/b^2}{ (2B^2/b^2 + 1) }
l_\mu l_\nu \; , 
\label{effective-metric} 
\ee

which is the result that we were looking for. The terms additional to the 
background metric $g_{\mu \nu}$ characterize any effective metric.

\subsection{Pagels-Tomboulis Abelian theory}

In 1978, the Pagels-Tomboulis nonlinear Lagrangian for electrodynamics 
appeared as an effective model of an Abelian theory introduced to describe  
a perturbative gluodynamics model. It was intended to investigate the non 
trivial aspects of quantum-chromodynamics (QCD ) like the asymptotic 
freedom and quark confinement \cite{pagels-tomboulis1978}. In fact, 
Pagels and Tomboulis argued that:

``{\textit{since in asymptotically 
free Yang-Mills theories the quantum ground state is not controlled 
by perturbation theory, there is no a priori reason to believe that 
individual orbits corresponding to minima of the classical action 
dominate the Euclidean functional integral}}. ''

In view of this drawback, 
of the at the time understanding of ground states in quantum theory,
they decided to examine and classify the vacua of the quantum gauge 
theory. To this goal, they introduced an effective action in which 
the gauge field coupling constant $g$ is replaced by the effective 
coupling $\bar{g}(t)\cdot T = {\rm ln} \left[\frac{F^a_{\mu \nu}F^{a
\, \, \mu \nu}}{\mu^4}\right]$. The vacua of this model correspond to 
paramagnetism and perfect paramagnetism, for which the gauge field is 
$F^a_{\mu \nu}=0$, and ferromagnetism, for which $F^a_{\mu \nu}F^{a\, 
\, \mu \nu} = \lambda^2$, which implies the occurrence of spontaneous 
magnetization of the vacuum. \footnote{This is the imprint that
such theory describes nonlinear electrodynamics.} 
They also found no evidence for instanton 
solutions to the quantum effective action. They solved the  equations 
for a point classical source of color spin, which indicates that in the 
limit of spontaneous magnetization the infrared energy of the field 
becomes linearly divergent. This leads to bag formation, and to an 
electric Meissner effect confining the bag contents. 

This effective model for the low energy (3+1) QCD reduces, in the Abelian 
sector, to a nonlinear theory of electrodynamics whose density Lagrangian 
$L(X, Y)$ is a functional of the invariants $X=F_{\mu\nu}F^{\mu\nu}$ and 
their dual $Y= (F_{\mu\nu}F^{\mu\nu})^\star$, having their equations of motion 
given by

\begin{equation}\label{eqsofmotion}
\nabla_\mu\left(-L_X F^{\mu\nu} - L_Y {^*}F^{\mu\nu}\right)=0\,,
\end{equation}

where $L_X=\partial L/\partial X$ and $L_Y=\partial L/\partial Y$. This 
equation is supplemented by the Faraday equation, i. e., the electromagnetic 
field tensor cyclic identity (which remains unchanged)

\begin{equation}\label{bianchi}
\nabla_\mu F_{\nu\lambda}+ \nabla_\nu F_{\lambda\mu}+ \nabla_\lambda
F_{\mu\nu}=0\,.
\end{equation}

In the case of a simple dependence on $X$, the equations of motion turn 
out to be \cite{kunze2008} (here we put $C=0$ and $4\gamma=-(\Lambda^8)^{
(\delta-1)/2}$ in the original Lagrangian given in  \cite{pagels-tomboulis1978})

\begin{equation}
L_{\delta}=-\frac{1}{4}\left(\frac{X^2}{\Lambda^8}\right)^{(\delta-1)/2} 
X\, ,
\label{pagels-tumb1978}
\end{equation}

where $\delta$ is an dimensionless parameter and $[\Lambda] = (an energy scale)$. 
The value $\delta=1$ yields the standard Maxwell electrodynamics. 

The energy-momentum tensor for this Lagrangian $L(X)$ can be computed by following the standard recipe, which then gives

\be
T_{\mu \nu} = \frac{1}{4 \pi} \left(L_X g^{ab} F_{\mu a}F_{b\nu} + g_{\mu \nu}L\right)
\ee

while its trace turns out to be 

\begin{equation}
 T=-\frac{1-\delta}{\pi}\left(\frac{X^2}{\Lambda^8}\right)^{(\delta-1)/2}X\,.
 \end{equation}

It can be shown \cite{kunze2008} that the positivity of the
$T_0^0\equiv \rho$ component implies that $\delta \geq 1/2$.

The Lagrangian (\ref{pagels-tumb1978}) has been studied by 
\cite{kunze2008} for explaining the amplification of the primordial
magnetic field in the Universe, being the analysis focused on three 
different regimes: 1) $B^2 \gg E^2$, 2) $B^2 \simeq
{\cal O}(E^2)$, 3) $E^2 \ll B^2$. It has also been used by 
\cite{herman-lambiase09} to discuss both the origin 
of the baryon asymmetry in the universe and the origin of primordial
magnetic fields. More recently it has also been discussed in the review on " Primordial magneto-genesis"
by \cite{kandus2010}.

Because the equation of motion (29) above, exhibits similar mathematical aspect as eq. (35) 
(reproduced in the Section), it appears clear that the Pagels and Tomboulis Lagrangian (31) 
leads also to an effective metric identical to that one given in equation (40), below.

\subsection{Novello-P\'erez Bergliaffa-Salim NLED}

More recently, \cite{NBS2004} (NPS)  revisited the several general 
properties of nonlinear electrodynamics  by 
assuming that the action for the electromagnetic field is that of Maxwell 
with an extra term, namely\footnote{Notice that this Lagrangian is gauge 
invariant, and that hence charge conservation is guaranteed in this theory.}

\be
S = \int \sqrt{-g} \left( - \frac F 4 + \frac \gamma F \right) d^4x \; ,
\label{action}
\ee

where $F\equiv F_{\mu\nu}F^{\mu\nu}$. 

Physical motivations for bringing in this theory have been provided in 
\cite{NBS2004}. Besides of those arguments, an equally unavoidable 
motivation comes from the introduction in the 1920's of both the Heisenberg-Euler and 
Born-Infeld nonlinear electrodynamics discussed above, which are valid in 
the regime of extremely high magnetic field strengths, i.e. near the Schwinger's 
limit. Both theories have been extensively investigated in the literature (see for 
instance \cite{herman-salim2004a, hermansalim2004b, NLED-REFERENCES1} 
and 
the long list of references therein).  Since in nature non only such very strong 
magnetic fields exist, then it appears to be promising to investigate also those 
super weak field frontiers. From the conceptual point of view, this 
phenomenological action has the advantage that it involves only the 
electromagnetic field, and does not invoke entities that have not been 
observed (like scalar fields) and/or speculative ideas (like higher-dimensions
and brane worlds). 

At first, one notices that for high values of the field $F$, the dynamics resembles 
Maxwell's one except for small corrections associate to the parameter $\gamma$, while 
at low strengths of $F$ it is the $1/F$ term that dominates. (Clearly, this 
term should dramatically affect, for instance, the photon-$\vec{B}$ field interaction in intergalactic 
space, which is relevant to understand the solution to the Pioneer anomaly  using NLED.). The consistency 
of this theory with observations, including the recovery of the 
well-stablished Coulomb law, was shown in \cite{NBS2004} using the cosmic microwave 
radiation bound, and also after discussing the anomaly in the dynamics of Pioneer 10 
spacecraft \cite{nos2006}. Both analysis provide small enough values for the coupling
constant $\gamma$ \cite{herman2010}. 

\subsubsection{Photon dynamics in NPS NLED: Effective geometry}

Next we investigate the effects of nonlinearities in the evolution of EM waves in the vacuum 
permeated by background $\vec{B}$-fields. An EM wave is described onwards as the surface of 
discontinuity of the EM field. Extremizing the Lagrangian $L(F)$, with $F(A_\mu)$, with respect 
to the potentials $A_{\mu}$ yields the following field equation \cite{plebanski70}

\be 
\nabla_{\nu} (L_{F} F^{\mu\nu} ) = 0
\label{eq60} , 
\ee

where $\nabla_\nu $ defines the covariant derivative. Besides this, we have 
the EM field cyclic identity

\be
\nabla_{\nu} F^{*\mu\nu} = 0 \hskip 0.3 truecm \Leftrightarrow 
\hskip 0.3 truecm F_{\mu\nu|\alpha} + F_{\alpha\mu|\nu} + 
F_{\nu\alpha|\mu} = 0\; . 
\label{eq62}
\ee

Taking the discontinuities of the field Eq.(\ref{eq60}) one gets 
(all the definitions introduced here are given in \cite{hadamard1903}) \footnote{Following 
Hadamard's method \cite{hadamard-method}, the surface of discontinuity of the EM field is 
denoted by $\Sigma$. 
The field is continuous when crossing $\Sigma$, while its first derivative presents a finite 
discontinuity. These properties are specified as follows: $\left[F_{\mu \nu} \right]_{\Sigma} 
= 0\; ,$ \hskip 0.3 truecm $\left[F_{\mu\nu|\lambda}\right]_{\Sigma} = f_{\mu\nu} k_\lambda\; 
\protect \label{eq14} \;$, where the symbol $\left[F_{\mu \nu}\right]_{\Sigma} = \lim_{\delta 
\to 0^+} \left(J|_{\Sigma + \delta}-J|_{\Sigma - \delta}\right)$ represents the discontinuity 
of the arbitrary function $J$ through the surface $\Sigma$. The tensor $f_{\mu\nu}$ is called 
the discontinuity of the field,  $k_{\lambda} = \partial_{\lambda} \Sigma $ is the propagation 
vector, and the symbols "$_|$" and "$_{||}$" stand for partial and covariant derivatives.}

\be 
L_{F} f^{\; \; \mu}_{\lambda} k^\lambda + 2L_{FF}F^{\alpha\beta} 
f_{\alpha\beta} F^{\mu\lambda} k_{\lambda} = 0 \; ,
\label{j1} 
\ee

which together with the discontinuity of the Bianchi identity yields

\be
f_{\alpha\beta}k_{\gamma} + f_{\gamma\alpha}k_{\beta} + 
f_{\beta \gamma} k_{\alpha} = 0\; .
\ee

A scalar relation can be obtained if we contract this equation
with $ k^{\gamma}F^{\alpha\beta} \label{eq25}$, which yields

\be
(F^{\alpha\beta} f_{\alpha\beta} g^{\mu\nu} + 2F^{\mu\lambda} 
f_{\lambda}^{\; \; \nu})k_{\mu} k_{\nu}=0 \; .
\label{j2}
\ee

It is straightforward to see that here we find two distinct solutions: a) 
when $F^{\alpha\beta} f_{\alpha\beta}=0$, case in which such mode propagates 
along standard null geodesics, and b) when $F^{\alpha\beta} f_{\alpha\beta}
=\chi$. In the case a) it is important to notice that in the absence of 
charge currents, this discontinuity describe the propagation of the wave front 
as determined by the field equation (\ref{eq60}), above. Thence, following \cite{lichnerowicz1962} the quantity $f^{\alpha\beta}$ can be decomposed in terms 
of the propagation vector $k_\alpha$ and a space-like vector $a_\beta$ (orthogonal 
to $k_\alpha$) that describes the wave polarization. Thus, only the light-ray having 
polarization and direction of propagation such that $F^{\alpha\beta} k_\alpha 
a_\beta = 0$ will follow geodesics in $g_{\mu \nu}$. Any other light-ray will 
propagate on the effective metric (\ref{eq63}). Meanwhile, in this last case, 
we obtain from equations (\ref{j1}) and (\ref{j2}) the propagation equation for 
the field discontinuities being given by \cite{Novello-etal2000}

\be
\underbrace{ \left(g^{\mu\nu} - 4\frac{L_{FF}}{L_{F}} F^{\mu\alpha} 
F_{\alpha}^{\; \; \nu}\right) }_{\rm effective \; metric} k_{\mu} k_{\nu} = 0 \; .
\label{eq63}
\ee

This equation proves that photons propagate following a geodesic that is
not that one on the background space-time, $g^{\mu\nu}$, but rather they 
follow the {\sl effective metric } given by Eq.(\ref{eq63}), which depends 
on the background field $F^{\mu\alpha}$, i. e., on the $\vec{B}$-field. 


\section{Understanding the Pioneer anomaly within NLED}

\subsection{Astrodynamics of Pioneer 10 and 11: Input facts}

As pointed out above, since 1998 the JPL group  have continuously reported 
an anomalous frequency shift derived from about ten years study of radio-metric 
data from Pioneer 10: 03/01/1987-22/07/1998 \cite{Anderson-etal1998}, 
Pioneer 11: 05/01/1987-01/10/1990 \cite{Anderson-etal1995}. The group has 
also found a similar feature in the data from  of Ulysses and Galileo spacecrafts 
\cite{Anderson-etal1998, Anderson-etal2002}.  The observed effect mimics a 
constant sunward acceleration acting on the spacecraft with magnitude

\be
a_{P} = (8.74 \pm 1.33) \times 10^{-8}~\rm cm~s^{-2}
\ee

and a steady frequency ($\nu$) drift 

\be
\frac{d \Delta \nu}{dt} \simeq 6 \times 10^{-9}~\rm Hz/s
\ee

which equates to a "clock acceleration": 

\be
\frac{d\Delta \nu}{dt} = \frac{a_{P}}{c}~\nu 
\label{freq.-drift} \,\, ,
\ee

where $c$ represents tha speed of light in a vacuum, and $t$ is the one way signal travel time. An
independent analysis of the radio-metric Doppler tracking data from the Pioneer 
10 spacecraft for the period 1987 - 1994 confirms the previous observations 
\cite{Markwardt2002}. In addition, by removing the spin-rate change contribution yields an 
apparent anomalous acceleration $a_{P} = (7.84 \pm 0.01) \times$ ~$10^{-8}$ cm~
s$^{-2}$, of the same amount for both Pioneer 10/11 \cite{Anderson-etal2002,
Abramovici-Vager86}. Besides, it has 
been noted that the magnitude of $a_{P}$ compares nicely to $cH_{0}$, where 
$H_{0}$ is the Hubble parameter today.

As stressed above, unlike other spacecrafts like the Voyagers and Cassini which 
are three-axis stabilized (hence, not well-suited for a precise reconstitution of 
trajectory because of numerous attitude controls), the Pioneer 10/11, Ulysses 
and the by-now destroyed Galileo are attitude-stabilized by spinning about an 
axis (parallel to the axis of the high-gain antenna) which permits precise 
acceleration estimations to the level of $10^{-8}$~cm~s$^{-2}$ (single 
measurement accuracy averaged over $5$ days). Besides, because of the 
proximity of Ulysses and Galileo to the Sun, the data from both spacecrafts 
were strongly correlated to the solar radiation pressure unlike the data from 
the remote Pioneer 10/11. Let us point out that the motions of the four 
spacecrafts are modelled by general relativistic equations (see \cite{Anderson-etal2002}, 
section $IV$) including the perturbations from heavenly bodies as small as 
the large main-belt asteroids (the Sun, the Moon and the nine planets are 
treated as point masses). Nonetheless, the observed frequency shift remains 
unexplained \cite{turyshev2010}. 

Thenceforth, several proposals for dedicated missions to test the 
Pioneer anomaly are now under consideration \cite{proposals-pioneer2005},  
in  virtue of the dramatic implications of the Pioneer puzzle for the 
understanding of gravity.\\

\subsection{What has been done by other researchers}

In search for a possible origin of the anomalous blueshift, a number of
gravitational and non-gravitational potential causes have been ruled out 
by \cite{Anderson-etal2002}. According to the authors, 
none of these effects may explain $a_{P}$ and some are $3$ orders of 
magnitude or more too small. The addition of a Yukawa force to the 
Newtonian law does not work ease. An additional acceleration is predicted 
by taking into account the Solar quadrupole moment
\cite{Mbelek-Michalski2004}. Although this entails a blueshift, 
it decreases like the inverse of the power four of the heliocentric radius, 
being of the order of $a_{P}$ only below $2.1$~AU.

Meanwhile, the claim that the Modified Newtonian Dynamics (MOND) may
explain $a_{P}$ in the strongly Newtonian limit of MOND \cite{Quevedo2005,
Milgron2001, Milgron2002} is not
obvious at all. First, the fits to the rotational curves of spiral galaxies 
yield for the MOND acceleration constant $a_{0}$ a value eight times smaller 
than $cH_{0}$ \cite{Quevedo2005}. Second, the gravitational pulling of the Sun up to 
$100$ AU still yields an acceleration greater than $a_{0}$ by at least three 
orders of magnitude, equating $a_{0}$ only at about $3000$ AU. Hence, Newtonian
dynamics up to general relativity corrections should apply to the spacecrafts.
Otherwise, one would be inclined to conclude that MOND is ruled out by a 
laboratory experiment \cite{Milgron2001, Milgron2002}. Now, any true Doppler
shift would involve 
an accompanying acceleration, which would be in conflict with both the 
motions of planets and long-period comets \cite{Anderson-etal1995, 
iorio2006a, iorio2006b, iorio2006c}.


Heretofore what we have learnt is that based on Einstein-Maxwell equations, 
the only other photon frequency shift that can be misinterpreted, at the 
solar system scale, with the Doppler shift is the gravitational frequency 
shift. In the weak field and low velocity limit, this would involve a time 
dependent gravitational potential instead of a spatial dependent one. Such 
proposals invoking the dark energy as the source of the time dependent 
gravitational potential have been suggested \cite{iorio2006a, iorio2006b, 
iorio2006c, tangen2006}. However, 
quintessence, like other fundamental scalar fields, has not yet been 
observed. 

In summary, prosaic explanations, non-gravitational forces and modified 
dynamics or new interaction (long or short range) force terms do not work
\cite{Mbelek-Michalski2004, Quevedo2005, Milgron2001, Milgron2002,
Ranada2003, Ranada2005}. Gravitational origin of the anomaly is rouled out 
by the precision of the 
planetary ephemeris (see \cite{Anderson-etal1998}, 
\cite{iorio2006a, iorio2006b, iorio2006c},  
and others \cite{tangen2006}) and the known bounds on dark matter within 
the orbital radius of Uranus or Neptune \cite{Ranada2003, Ranada2005, 
Whitmire-Matese2003}.

\subsection{What we are proposing to tackle the Pioneer anomaly}

By gathering together all the arguments reviewed above, one is led to the 
conclusion that the Pioneer anomaly does not seem to be related to the 
gravitational interaction \cite{Anderson-etal1998, iorio2006a, iorio2006b,
iorio2006c, tangen2006}. If this is the 
case, what other of the currently known interactions in nature could afford 
a consistent understanding for the radio-metric Doppler tracking data from 
Pioneer spacecrafts? 

The right answer could be related to the fact that there are only two 
long range interactions known today: Gravity and electromagnetism. 
Therefore, what remains is the EM 
sector.\footnote{Non-metric fields can also be regarded as gravitational 
fields and there is a lot of space for speculation.} Meanwhile, the 
possibility of an interaction of the EM signal with the solar wind 
leading to a change of the frequency of the EM signal is now rouled 
out (see \cite{Anderson-etal2002}). 

Indeed, it appears to be unescapable to conclude that what we are observing 
(measuring through the receivers) could be related to the equation of motion 
of the photon. In other words, the mounting evidence seems to converge to what 
could be happening to the photon during its propagation through the interplanetary 
space from the Pioneer 10/11 antennas to the receivers on Earth. 

It is timely, then, to recall that classical (Maxwell theory) or quantized (QED) 
linear electrodynamics does not allow for a change of the frequency of a photon 
during its propagation in a linear medium without invoking diffusion due 
to the interaction with the surrounding matter (hence a smear out of the 
image of the source). Moreover, for such a phenomenon to occur, one needs 
to consider a general and non trivial Lagrangian density $L = L(F)$ for which its second 
derivative w.r.t. $F$: $d^2L/dF^2 = L_{FF} \neq 0$. Therefore, the Pioneer 
anomaly, if not an artifact, may be a result of NLED as we show below. 
Indeed, relation (\ref{freq.-drift}) above translates, in covariant 
notation, into 

\be
\frac{dx^{\nu}}{d{ l}}~\nabla_{\nu}~k^{\mu} = \frac{a_{P}}{c^2}~
k^{\mu} \, , 
\ee

where ${ l}$ is some affine parameter 
along a ray defined by $k^{\mu} = \frac{dx^{\mu}}{d{ l}}$ 
(see \cite{Fujii06}). The latter equation departs 
from the classical electrodynamics one (see \cite{LL1970}, 
section 87)

\be
\frac{dx^{\nu}}{d{ l}}~\nabla_{\nu}
~k^{\mu} = 0
\ee

and suggests the occurrence of the NLED effect dubbed photon acceleration. 

The concept of photon acceleration, which follows 
from the description of photon propagation in NLED, was discussed 
by \cite{Novello-salim2001}, see also the book by 
\cite{mendonca-book}. Next we explain why the anomaly shows up in 
some situations and not others. (For experimental tests of NLED and 
further theoretical predictions see \cite{NLED-REFERENCES1, 
NLED-REFERENCESA, NLED-REFERENCES0, NLED-REFERENCES2, 
NLED-REFERENCES3}).

Therefore, the alternative that the Pioneer anomaly is not consequence 
of an actual change in the spacecraft velocity (see \cite{Anderson-etal2002}, 
Section X) deserves to be investigated. Indeed, a direct interpretation of the 
observational data from the spacecrafts 
implies merely an anomalous time-dependent blueshift of the photons of the 
communication signals. On the other hand, in using a time dependent potential 
\cite{iorio2006a, iorio2006b, iorio2006c, tangen2006} to explain the Pioneer 
10/11 data one may be pointing out to 
the need of an \textit{effective metric} for the photons. In fact, what is 
needed is just a time variation of the $4$-momentum of the photon along 
its path. Thus the atomic energy levels would not be affected. Rather, only the 
motion of the photon being concerned.

\subsection{NLED at all distance scales: From cosmology down to astrodynamics in the Solar System} 

Upon the collection of arguments presented above, it appears that all these 
requirements are achieved by considering that 
NLED is based on a Lagrangian density $L(F)$ which includes terms depending
nonlinearly on the invariant $F = F_{\mu\nu}~F^{\mu\nu}$, with $F = 2 (B^2 
c^2 - E^2)$ \cite{Novello-etal2000, Novello-salim2001, plebanski70}, instead 
of the usual Lagrangian density $L = - \frac{1}{4} F$ of  classical 
electromagnetism in a vacuum. As stated above, we shall explore the effects 
of nonlinearities in the evolution of EM waves, which are envisioned onwards 
as the surface of discontinuity of the EM field.  Therefore, as shown above, 
by extremizing the Lagrangian with respect to the potentials $A_{\mu}$ one 
obtains the EM field equation of motion  
\cite{plebanski70}\footnote{Next we show that the "acceleration" of photons
predicted by NLED may account for the anomalous blueshift indicated by the
Pioneer 10/11, Ulysses and Galileo spacecrafts.  This will manifest itself
as a new frequency shift for the EM waves, in addition to the Doppler shift 
(special relativity) and the gravitational and cosmological redshift (general 
relativity), when both of them apply.}  

\be
\nabla_{\nu} (L_{F} F^{\mu\nu} ) = 0
\label{eq60A} ,
\ee

in which $\nabla_\nu $ represents the covariant derivative, and $L_F = 
dL/dF$. 

Recalling the discussion above, the dynamics of the photon propagation 
follows the equation 

\be
\left(g^{\mu\nu} - 4\frac{L_{FF}}{L_{F}} F^{\mu\alpha}
F_{\alpha}^{\, \, \nu}\right) k_{\mu}k_{\nu} = 0 \, .
\label{63A}
\ee

which exhibits the fundamental feature of NLED, i.e., the effective 
metric.

Then, by taking the derivative of the last expression, one arrives to

\be
k^{\nu} \nabla_\nu k_{\alpha} = 4 \left(\frac{L_{FF}}{L_{F}}F^{\mu\beta}
F_{\beta}^{\, \, \nu} ~k_{\mu} k_{\nu}\right)_{|\alpha}.
\label{knuknuA}
\ee

Eq.(\ref{knuknuA}) shows that the nonlinear Lagrangian introduces a term 
acting as a force accelerating the photon. This acceleration of any photon 
which is traversing over weak background electromagnetic fields in a vacuum 
is the new physical element that we argue hereafter would be responsible 
for the Pioneer anomaly.

\subsubsection{NLED photon acceleration: What Earth receivers are reading 
off radio signals from interplanetary spacecraft transponders - The case 
Pioneer anomaly} 

If NLED is to play a 
significant role at the macroscopic scale, this should occur at the 
intermediary scales of clusters of galaxies or the interclusters medium, 
wherein most observations show that the magnetic fields are almost uniform 
(and of the same order of magnitude\footnote{\cite{Fujii06}}), 
unlike the dipolar magnetic fields of the Sun and planets. However, galaxies 
are gravitationally bound systems, whereas the cosmic expansion is acting at 
the cluster of galaxies scale. Thus, the magnetic field (${\bf B}$) in clusters 
of galaxies (IGMF) depends on the cosmic time ($B = B_0 a^{-2}$). So, the 
${\bf B}$ that is relevant to this study is that of the local cluster of galaxies 
\cite{beck2000}. 
(As regard to the contribution of the CMB radiation see \cite{estimating-gamma}).\footnote{The 
interclusters magnetic field is in any case by far small ($10^{-9}$
~G) to add a measurable correction even to the cosmological redshift. 
As for the contribution of the cosmic microwave background (CMB), not only 
it is too weak but also, the CMB is pure radiation ($F = 0$), whereas we are 
interested in the case of a background magnetic field with no significant electric 
field counter-part, i.e., $E=0$.} Recently, \cite{Vallee02} has speculated 
that the $2~\mu$G magnetic field he has observed within the local supercluster 
of galaxies in cells of sizes of about $100$~kpc may extend all the way down to 
the Sun. We explore further this idea in the framework of NLED and show that it is 
capable to provide an explanation of the Pioneer anomaly from first principles.

Relation (\ref{eq63}) can be cast in the form 

\be
g_{\mu\nu} k^{\mu}k^{\nu} = 4\frac{L_{FF}}{L_{F}} b^2,
\label{64}
\ee

where $b^\mu = F^{\mu\nu} k_\nu$ and $b^2 = b^\mu b_\mu$. 

As $E=0$, one can write, 
after averaging over the angular-dependence \cite{BB1970}: 

\be
b^2 = - \frac{1}{2}||\vec{k}||^2 B^2 c^2 = - \frac{1}{4}||\vec{k}||^2 F \, 
\label{65}
\ee 

with $||\vec{k}|| = \omega/c = 2\pi \nu/c$. By inserting this relation in (\ref{64}) yields

\be
g_{\mu\nu} k^{\mu}k^{\nu} = - \frac{\omega^2}{c^2} \,F \, \frac{L_{FF}}{L_{F}}.
\label{66}
\ee

Taking the ($x^{\alpha}$) derivative of Eq.(\ref{66}) we obtain

\be
2 g_{\mu\nu} k^{\mu} (k^{\nu})_{|\alpha} \,+ \,k^{\mu} k^{\nu}
(g_{\mu\nu})_{|\alpha} = - \left(\frac{\omega^2}{c^2} \, F \, 
\frac{L_{FF}}{L_{F}}\right)_{|\alpha}.
\label{67}
\ee

The cosmological expansion will be represented by $g_{\mu\nu} = a^{2}(\eta) 
g_{\mu\nu}^{(local)} \; \label{68}$, with $a$ the scale factor, $\eta$ the
conformal time, and $g_{\mu\nu}^{(local)}$ the local metric. So, Eq.(\ref{67}) 
yields: 

\be
2 g_{\mu\nu} k^{\mu} (k^{\nu})_{|0} \,+ \,2 \left(\frac{\dot{a}}{a}\right) \,g_{\mu\nu}
k^{\mu} k^{\nu} = -\, \left(\frac{\omega^2}{c^2} \,F \, \frac{L_{FF}}{L_{F}}\right)_{|0}
\; , \label{69}
\ee

where the dot stands for partial derivative w.r.t. $\eta$. Using 
Eqs.(\ref{66}) and (\ref{69}) we obtain\footnote{By removing the 
NLED extra term from Eq.(\ref{64}), this reduces it to $g_{\mu\nu
}^{(local)} k^{\mu}k^{\nu} = 0$ so that the photons would just 
see the local background metric.}

\be
k_{\mu} (k^{\mu})_{|0} = \left(\frac{\dot{a}}{a}\right) \, \frac{\omega^2}
{c^2} \,F \, \frac{L_{FF}}{L_{F}} \,- \, \frac{1}{2} \, \left(\frac{\omega^2}{c^2} 
\,F\, \frac{L_{FF}}{L_{F}}\right)_{|0}.
\label{70}
\ee

Now, $\dot{F} = \,- \,4 \, \left(\frac{\dot{a}}{a}\right) \,F, \label{71}$ by recalling 
that $B^2 \propto a^{-4}$. Moreover, from the method of the effective metric, it can 
be shown that $k^{0}$ does not vary with time in the first order approximation 
unlike $||\vec{k}||$.\footnote{Given a background metric
$g_{\mu\nu}$, as a result of NLED effects photons follow geodesic 
paths with respect to the effective metric (or any one conformal to it)
$g^{(eff)}_{\mu\nu} = g_{\mu\nu} - 4\frac{L_{FF} }{L_{F} } F_{\mu}^{\, \, 
\alpha} F_{\alpha\nu}$ (see \cite{Novello-etal2000},\cite{Novello-salim2001}).
Thus, following the usual analysis on the gravitational frequency shift 
but with $g^{(eff)}_{\mu\nu}$ replacing $g_{\mu\nu}$, one gets $k^{0} c = 
{\omega}_{0}/\sqrt{g^{(eff)}_{00}}$ (see \cite{LL1970}, section 88), 
where ${\nu}_{0} = {\omega}_{0}/2\pi$ denotes the photon frequency in 
flat Minkowski spacetime. Thus, discarding the cosmological redshift 
(subsequent to the time dependence of the curvature), the variation of 
$k^{0}$ with time can be neglected in the first order approximation, 
since $F^{0\alpha} F_{\alpha}^{\, \, 0} =  F_{0}^{\, \, \alpha}
F_{\alpha 0} = 0$ in the case of a zero electric field.}  Hence 

\be
k_{\mu} (k^{\mu})_{|0} = - \,
\frac{\omega}{c} \, \left(\frac{\dot{\omega}}{c}\right)\; .
\label{72}
\ee 

By inserting relation (\ref{72}) in (\ref{70}), and then expanding 
and arranging, one finds

\be
\left(\frac{\dot{\nu}}{\nu}\right) = - \, \left(\frac{\dot{a}}{a}\right) \,\frac{Q + 2F Q_{F}}{1 
- Q}.
\label{73}
\ee

where we have set $Q = F \, \frac{L_{FF}}{L_{F}}$ and $Q_{F} = \partial
Q/\partial F$.

At present cosmological time ($t$), and for a duration very short as compared 
to the universe age, Eq.(\ref{73}) reduces to 

\be
\left(\frac{\dot{\nu}}{\nu}\right) \simeq - \, 
H_{0} \, \frac{Q + 2 F Q_{F}}{1 - Q} 
\label{74}
\ee

where ($\dot{\nu}$  is the photon frequency $t$-derivative). $\dot{\nu} \neq 0$ 
if and only if a) the NLED contribution is non-null, i.e., $L_{\rm FF} \neq 0 $, 
and b) $F$ depends on time.

\subsubsection{NLED {\textit{alla}} NPS as explanation of the Pioneer puzzle}

The explicit form of this general nonlinear Lagrangian (which simulates the 
effect of dark energy in.\cite{NBS2004}) reads

\be
L = - \frac{1}{4} F + \frac{\gamma}{F} \,  , \hskip 0.3
truecm {\rm or  } \hskip 0.3 truecm L = - \frac{1}{4} F +
\frac{\gamma_n}{F^n} \, ,
\label{12}
\ee

where $n$ is a strictly positive integer. From Eqs.(\ref{73},\ref{12}), 
the time variation of the photon frequency, due to interaction with very 
weak {\bf B}$(t)$ fields, reads

\begin{equation}
\left(\frac{\dot{\nu}}{\nu}\right) =  A_n \gamma_n \frac{4n\gamma_n - (2n+1)
F^{n+1} }{(F^{n+1} + 4 n \gamma_n) \left[F^{n+1} + 4n(n+2) \gamma_n\right]}.
\label{14}
\end{equation}

with $A_n = 4H_0 n(n+1)$. Notice that $\gamma_n$ should be negative in order
to guarantee that the Lagrangian is bound from below (see \cite{LL1970},
sections 27 and 93), $\gamma_n = - (B_n c)^{2(n+1)}$. Also, it is worth 
noticing that Eq.(\ref{14}) in the nearly-zero field limit ($B \rightarrow 0$) 
would reduce to

\begin{equation}
\left(\frac{\dot{\nu}}{\nu}\right) =  H_0 \left(\frac{n+1}{n+2}\right)\, , 
\label{16}
\end{equation}

which implies a blueshift.

\subsection{Discussion and conclusion}

We stress that the NLED is a universal theory for the electromagnetic field, 
with $\gamma_{n=1} = \gamma$ in Eq.(\ref{12}) being a universal constant. 
The value of $\gamma$ was fixed in \cite{NBS2004} by using the CMB 
constraint.\footnote{Nonetheless, we stress that a conclusive method 
of fixing $\gamma$ should benefit of a dedicated laboratory experiment, in
the same spirit that it was done, for instance, to fix the electron charge 
through Millikan's experiment.}  Indeed, in the standard model of cosmology 
the CMB is well described by Maxwell
theory, which is likely to give a good account of the magnetic fields in
galaxies too. Notwithstanding, the processes at the origin of both the seed 
magnetic
field in the interclusters medium and clusters of galaxies are not yet
clearly understood. Hence, if NLED is to play a significant role at the
macroscopic scale, this should occur at the intermediary scales of clusters
of galaxies or the interclusters medium. 

Then, considering the possibility that the NLED correction terms described 
above come into play at these scales, one gets the following ordering of 
strengths: $B_{\rm Universe} \ll B_{\rm Intercluster} \ll B_1 \lesssim 
B_{\rm Cluster} \lesssim B_{\rm Galaxy}$. Turning back to the Pioneer 
anomaly, a good accordance is obtained with $B_{\rm LSC} = 1.77~B_1$ 
(case $n=1$) \cite{estimating-gamma}. 
Thus, setting $B_1 = \frac{1}{c} |\gamma|^{1/4}$, one finds $B_1 = 0.008 \pm 0.002 
~\mu$G {\cite{estimating-gamma}}.\footnote{\label{B1-Value} 
A clean estimate of $B_1$ from our definition 
of $\gamma_n \equiv - (B_n c)^{2(n+1)}$, below Eq.(\ref{14}), and the one 
in \cite{NBS2004}: $\gamma \equiv - \hbar^{2} ~\mu^{8}$. On account 
that $a_{c} = (1 + z_{c})^{-1}$ and $\gamma = -  \hbar^{2} ~\mu^{8} = - (B_{1} 
c)^{4}$, Eq.(13) of \cite{NBS2004} rewrites $B_{1}^{4}/\mu_{0}~B_{0}^{2} 
= 1.40 \rho_{c} c^{2}$ (A1). Thus $a_{c} = (B_{0}^{4}c^{4}/3\hbar^{2}\mu^{8})^{1/8}$ 
yields $B_{1} = 3^{-1/4} (1 + z_{c})^{2} B_{0}$ (A2). Then, combining both 
relations (A1) and (A2) one gets : $B_{0} = 0.02~(1 + z_{c})^{-4}$~$\mu$G (A3),
$B_{1} = 0.016~(1 + z_{c})^{-2}$~$\mu$G (A4). Now, Riess etal. found evidence 
for a transition from cosmic deceleration to acceleration at redshift $z_{c} = 
0.46 ~\pm ~0.13$ [A. G. Riess etal., ApJ 607, 665 (2004)]. Inserting the latter 
figures in relations (A3) and (A4)  yields : $B_{0} = (0.005 ~\pm ~0.002)$~
$\mu$G (A5), $B_{1} = (0.008 ~\pm ~0.002)$~$\mu$G (A6). Since CMB is pure 
radiation (i. e., $E = B c$ not equal to zero on average), we consider that 
relations (A4) and (A6) give a better estimate of $B_{1}$ than the one put 
forward in \cite{NBS2004}.}

Thence, to compute the effect (drift) on the Pioneer communication signal 
frequencies (uplink and downlink), we need only to introduce the value of the 
strength of the local supercluster {\bf B}-field: $B_{\rm LSC} \sim 10^{-8} - 
10^{-7}$~G \cite{Blasi1999}. This involves $Q > 0$. But then, since relation 
(\ref{66}) implies $Q < 1$, and the NLED theory must be such that $L_{\rm F} 
< 0$ (hence, $F^{n+1} + 4n \gamma_n > 0$) for the energy density of the EM 
field to be positive definite [see \cite{Novello-etal2000}, appendix B], which 
entails {\bf B}$ > B_{1}$, one can verify that the equation (\ref{14}) implies 
a blueshift.

Meanwhile, the impressive accordance of the data from Voyager 1 and 2 
magnetometers 
with Parker's theory (see  Fig. 2) constraints $B_{\rm LSC}$ to be less 
than $0.022~\mu$G within the solar system up to the heliopause. Hence, 
we may conclude that $0.01~\mu$G $< B_{\rm LSC} < 0.022~\mu$G within 
the solar system. By recalling that the uplink frequency of Pioneer 10 and 
11 spacecrafts is $\nu = 2.2$~GHz, one obtains for the median value 
$B_{\rm LSC} = 0.018~\mu$G (both expressions are normalized by 
$\left[\frac{H_0}{70~\rm{km} ~{s^{-1} ~Mpc^{-1}}}\right]$).  Then
Eq.(\ref{16}) renders 

\begin{eqnarray}
\left(\frac{\dot{\nu}}{\nu}\right) = 2.8 \times 10^{-18} ~ {\rm s^{-1}}\, , 
\hskip 0.5 truecm {\rm or \, \, equivalently} \hskip 0.5 truecm 
\frac{d\Delta \nu}{dt} = 6 \times 10^{-9} \, \rm \frac{Hz}{s}\, ,
\label{dot-freq}
\end{eqnarray}

with $\Delta \nu$ being the frequency drift pointed out earlier.


\textbf{A digression on interplanetary magnetic field and NLED effects:}

 It has been pointed out that the strength of the IPMF could severly 
minimize the NLED effects { because it will overrun the interstellar 
or intergalactic magnetic fields at heliocentric distances}. Notwithstanding, 
the actual data from Voyager 1/2 spacecrafts of the IPMF average strength 
(see  Fig. 2) are both consistent with a non-zero local supercluster magnetic 
field (LSCMF) 
amounting up to $0.022~\mu$G \cite{voyager1,voyager2} (the accuracy 
of the measurements performed by Pioneer 10/11 magnetometers is at best $0.15~ 
\mu$G, and $0.022~\mu$G for the low field system of Voyager 1/2 magnetometers 
\cite{solar-wind}). Besides, it is just beyond the Saturn orbit, $\sim 10$ 
Astronomical Units (AU), that the anomaly begins to be clearly observed.  
Surprisingly, it is just after passing the Saturn orbit that the strength
of the magnetic field vehiculated by the solar wind gets down the strength 
associated to the insterstellar and intergalactic magnetic fields, as one can 
verify by perusing on Fig. 2 (see Refs.\cite{voyager1,voyager2}). Thus, since 
a magnetic field 
cannot shield (or block) any another magnetic field (the stronger field can only
reroute the weaker  field, otherwise it would violate Maxwell's laws), then 
it follows that the LSCMF has its magnetic influence extended upto nearly the 
location of the Saturn orbit, and in this way it forces the photons being emitted
by the Pioneer spacecrafts from larger heliocentric distances to get accelerated 
due to the NLED effects.  Besides, notice that \cite{mitra2003} also shows 
that the local cloud of interstellar gas in HII regions does not keep out the 
Galactic magnetic field.

``En passant'', we call to the reader's attention the fact that some workers 
in the field have claimed that the effect should 
have shown up already at the small distance corresponding to Mars, Jupiter or 
Saturn orbits, because of the high technology involved in the tracking of planet 
orbiting spacecrafts as Galileo and Cassini or the Mars' nonroving landers, 
which would allow to single out the anomaly at those heliocentric distances. 
However, as those spacecrafts are inside the region where the solar wind dominates, 
this definitely precludes the NLED photon acceleration effect to show up at
those distances since the much higher magnetic field there would introduce 
a negligible NLED effect, and as stated below the solar pressure influence on
the signal frequency is still large. 

On the other hand, although one 
can use the time of flight of photons during tracking of planets with orbiting 
spacecrafts (by combining range and Doppler data over a spacecraft orbit) to 
tightly determine the range from Earth to that given planet's center of mass, 
the impediment to single out the radio-signal frequency shift remains the same 
pictured above: from one side the strength of both the host planet magnetic 
field and the solar magnetic field at those distances are  still large, what blocks 
the nonlinear action of the LSCMF, and from the 
large up to 20 AU so as to allow the show up of the NLED frequency shift which 
is much smaller.  Moreover, within a heliocentric distance $\sim$ 100~AU the 
IPMF keeps stronger. 
Thus, it reduces for all practical purposes the IPMF contribution to the effects 
of NLED (see further arguments from our direct estimate of $B_1$ in Footnote 
\ref{B1-Value} above, and in \cite{estimating-gamma}), leaving room 
for the sole contribution of the residual IGMF in the solar system.

Finally, the new frequency shift that is predicted by NLED is not seen yet in the 
laboratory { because of some of the following reasons: a) the most important, 
the strength of the Earth magnetic field is much larger than the one required in 
the NLED explanation of the anomaly for the effect to show up, and b) } the 
coherence time $\tau = 1/\Delta\nu$ of EM waves in 
present atomic clocks (frequency width $\Delta\nu > 0.01$ Hz, or otherwise 
stated $c\tau < 0.2$ AU) is too short as compared to the time of flight of 
photons from Pioneer 10/11 spacecrafts past 20 AU. Nonetheless, if the 
conditions demanded by our model were satisfied this effect will certainly 
be disentangled in a dedicated experiment { where, for instance, the Earth 
magnetic field is kept outside the case containing an experimental set up
where a very weak magnetic field is maintained inside, a source of photons 
set to travel and a receiver  data-collecting.} 

\section{Acknowledgements}

\begin{figure}[p]	
\centering
\subfigure[Subfigure 1]{\includegraphics[width=6.6cm]{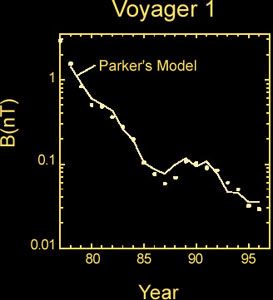}}
\subfigure[Subfigure 2]{\includegraphics[width=6.6cm]{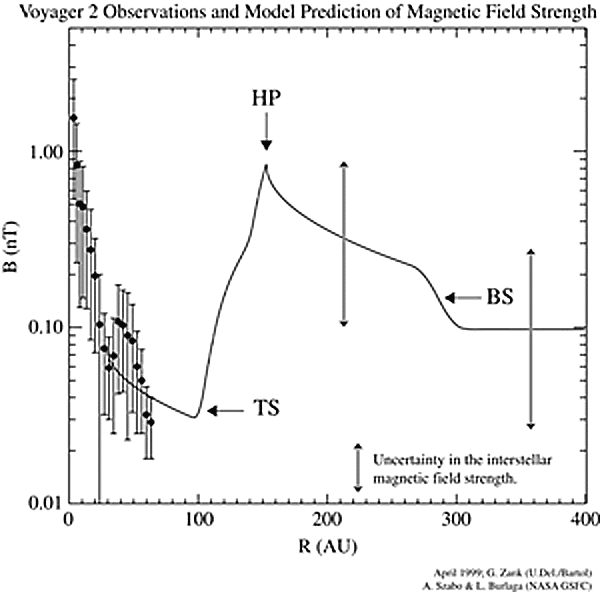}}
\caption{Data from Voyager 1/2 spacecrafts of the interplanetary magnetic field (IPMF) average strength. Subfigure 1 presents measurements by Voyager 1 of the strength of the IPMF [nT] as a function of time [Yr]. The continuous line represents the predictions of Parker's model, and the dots the Voyager 1 data. Subfigure 2 shows the dependence with the distance [AU] of the IPMF [nT], as detected by Voyager 2, against the theoretical prediction. Credit: http://spacephysics.ucr.edu/images/swq2-04.jpg, and http://interstellar.jpl.nasa.gov/interstellar/probe/interstellar/images/ 07BupstrmSuess-25lg.gif}
\label{fig:Sub_1}
\end{figure}

The author would like to sincerely thank Jean-Paul Mbelek 
(Service d'Astrophysique, C. E. Saclay, 91191 Gif-sur-Yvette Cedex, France), 
M\'ario Novello (Instituto de Cosmologia, Relatividade e Astrof\'{\i}sica 
(ICRA-BR), Centro Brasileiro de Pesquisas F\'{\i}sicas (CBPF)),  and Jos\'e 
Martins Salim (Instituto de Cosmologia, Relatividade e Astrof\'{\i}sica
(ICRA-BR), Centro Brasileiro de Pesquisas F\'{\i}sicas (CBPF)) for the former
collaboration in this direction of research. Also Prof. Remo Ruffini, Director
General of the International Center for Relativistic Astrophysics Network 
(ICRANet), and the International Coordinating Centre, Pescara, Italy are 
thanked for the hospitality given during the final preparation of this work 
(September 2010).





\section{APPENDIX: The method of effective geometry}

Following Hadamard (1903), the surface of discontinuity\footnote{ Of course, 
the entire discussion onwards could alternatively be rephrased in terms of 
concepts more familiar to the astronomy community as that of light rays used 
for describing the propagation of electromagnetic waves in the  geometric optics 
approximation.} of the EM field is denoted by $\Sigma$. The field is continuous
when crossing $\Sigma$, while its first derivative presents a finite
discontinuity. These properties are specified as follows

\begin{eqnarray}
[B^\mu]_{\Sigma} & = & 0, \hskip 0.3 truecm \left[\partial_\alpha 
B^\mu\right]_{\Sigma} = b^\mu k_\alpha, \hskip 0.3 truecm 
\left[\partial_\alpha E^\mu \right]_{\Sigma} = e^\mu k_\alpha \nonumber \\ \nonumber \\
\left[F_{\mu\nu}\right]_{\Sigma} & = & 0\; ,  \hskip 0.3 truecm \left[F_{
\mu\nu|\lambda}\right]_{\Sigma} = f_{\mu\nu}k_\lambda\; 
\protect\label{eq14} \; , 
\end{eqnarray}

where the symbol 

\begin{equation} 
\left[F_{\mu\nu}\right]_{\Sigma} \equiv {\rm lim}_{\delta \rightarrow 0^+}
(J|_{\Sigma+\delta}-J|_{\Sigma-\delta}) \; \label{discontinuity} \; 
\end{equation}

represents the discontinuity of the arbitrary function $J$ through the
surface $\Sigma$. The tensor $f_{\mu\nu}$ is called the discontinuity
of the field,  and $k_{\lambda} = \partial_{\lambda} \Sigma $ is the
propagation vector. In Eq. (\ref{eq14}),  the symbol ``$_|$'' stands 
for partial derivative; 

Here-after we want to investigate the effects of nonlinearities of
very strong magnetic fields in the evolution of electromagnetic waves
described onwards as the surface of discontinuity of the electromagnetic
field (represented here-to-fore by $F_{\mu\nu}$). For this reason 
we will restrict our analisys  to the simple class of gauge invariant 
Lagrangians defined by $L = L(F) \label{eq1}$. From the least action 
principle we obtain the following field equation

 \be
 \left(L_F F^{\mu\nu}\right)_{||\mu} = 0 \; .\label{eq4} \
 \ee

Applying the Hadamard conditions (\ref{eq14}) and (\ref{discontinuity}) 
to the discontinuity of the field in Eq.(\ref{eq4}) we obtain

\be 
L_F f^{\mu\nu}k_\nu+2L_{FF} \xi F^{\mu\nu}k_\nu = 0\; ,\label{eq5} 
\ee 

where $\xi$ is defined by $\xi\doteq F^{\alpha\delta}f_{\alpha\delta}$. 
Both, the discontinuity conditions and the electromagnetic field tensor 
cyclic identity lead to the following dynamical relation

 \be
f_{\mu\nu}k_{\lambda}+f_{\nu\lambda}k_{\mu}+f_{\lambda\mu}k_{\nu}=0
\label{eq6} \; .
\ee 

In the particular case of a polarization such that $\xi=0$, it follows 
from Eq.(\ref{eq4}) that $f^{\mu\nu}k_\nu = 0 $. Thus using 
this result, and multiplying Eq.(\ref{eq6}) by $k^\lambda$ we obtain

 \be
f_{\mu\nu} k^{\mu} k^{\nu} = 0\; . \label{eq7}
 \ee

This equation states that for this particular polarization the
discontinuity propagates with the metric $f_{\mu\nu}$ of the 
background space-time. For the general case, when $\xi \neq 0$, 
we multiply Eq.(\ref{eq6}) by $k_\alpha g^{\alpha\lambda} F^{\mu\nu}$ 
to obtain

\be 
\xi k_\nu k_\mu g^{\mu\nu}+2F^{\mu\nu}f_\nu^\lambda k_\lambda k_\mu = 0 \; . 
\ee

From this relation and  Eq.(\ref{eq5}) we obtain the propagation
law for the field discontinuities, in this case given as 

\be 
\left(L_F g^{\mu\nu} - 4L_{FF} F_{\alpha}^{\mu} F^{\alpha\nu} \right)
k_\mu k_\nu = 0 \; ,  \label{ef-metric}
\ee 

where $ F^{\mu }_{{\;} {\;} {\;} \alpha}  F^{\alpha \nu}  = - B^2 h^{\mu \nu} 
- B^\mu B^\nu \label{F-contracted}$. Eq.(\ref{ef-metric}) allows  to interpret 
the term inside the parenthesis multiplying $k^\mu k^\nu $ as an effective 
geometry 

\be 
g^{\mu\nu}_{\rm eff} = L_Fg^{\mu\nu}-4L_{FF} F^{\mu}_{\alpha} 
F^{\alpha\nu}\; . \label{eff-metric} 
\ee 

Hence, one concludes that the discontinuities will follow geodesics in this 
effective metric.

\section{APPENDIX-1: NPS theory applied to cosmology}

To discuss the evolution of a universe model driven by the NPS NLED, the electromagnetic (
EM) field described by Eq.(\ref{action}) can be taken as 
source in Einstein equations to obtain a toy model for the evolution of the universe which 
displays accelerate expansion. Such phase of acceleration runs into action when the nonlinear 
EM term takes over the term describing other matter fields. This NLED theory yields ordinary 
radiation plus a dark energy component with $w < -1$ (phantom-like dynamics). Introducing the 
notation\footnote{Due to the isotropy 
of the spatial sections of the Friedman-Robertson-Walker (FRW) model, an average procedure 
is needed if electromagnetic fields are to act as a source of gravity \cite{tolman-ehrenfest}. 
Thus a volumetric spatial average of a quantity $X$ at the time $t$ by $\langle X \rangle_{|_V} 
\equiv \lim_{V\rightarrow V_0} \frac 1 V \int X \sqrt{-g}\;d^3x$, where $V = \int \sqrt{-g} 
\;d^3x$ , and $V_0$ is a sufficiently large time-dependent three-volume. (Here the metric sign 
convention $(+---)$ applies).}, the EM field can act as a source for the FRW model 
if $\langle E_i \rangle_{|_V} =0, \; \langle B_i \rangle_{|_V} =0,\; \langle {E_i B_j} \rangle_
{|_V} = 0$, $ \langle {E_iE_j} \rangle_{|_V} = - \frac 1 3 E^2 g_{ij}$, and $\; \langle {B_iB_j} 
\rangle_{|_V} = -\frac 1 3 B^2 g_{ij}$.\footnote{Let us remark that since we are assuming that 
$\langle {B}_i \rangle_{|_V} = 0$, the background magnetic fields induce no directional effects 
in the sky, in accordance with the symmetries of the standard cosmological model.} When these 
conditions are fulfilled, a general nonlinear Lagrangian $L(F)$ yields the energy-momentum tensor
($L_F = {dL}/{dF}, \;\; L_{FF} = {d^2L}/{dF^2}$)\footnote{Under the same 
assumptions, the EM field associate to Maxwell Lagrangian generates the 
stress-energy tensor defined by Eq.(\ref{tmunu}) but now $ \rho = 3 p = 
\frac{1}{2} (E^2 + B^2)$.}  

\begin{eqnarray}
\nonumber \\
& \langle {T}_{\mu\nu} \rangle_{|_V}  =  (\rho + p) v_\mu v_\nu - p\; 
g_{\mu\nu} , & \label{tmunu} \\ \nonumber \\
& \rho  =  -L - 4E^2 L_F , \;\;\;\;\; p =  L + \frac 4 3 (E^2-2B^2) L_F , 
& \nonumber \\ \nonumber \\ \nonumber
\end{eqnarray}

Hence, when there is only a magnetic field, the fluid can be thought of as 
composed of a collection of non-interacting fluids indexed by $k$, each of 
which obeys the equation of state $p_k = \left( \frac{4k}{3} - 1 \right) \rho_k 
\label{partialp}$, composed of ordinary radiation with $p_{1}= \frac 1 3\; 
\rho_{1}$ and of another fluid with equation of state $p_{2} = -\frac 7 3 
\;\rho_{2}$. It is precisely this component with negative pressure that may 
drive accelerate expansion through Friedmann equations, as was shown in 
\cite{NBS2004}.


\end{document}